\documentstyle[aps,pra,eqsecnum,preprint,epsfig]{revtex}

\tighten

\begin{document}

\narrowtext

\title{Fidelity trade-off for finite ensembles of identically
prepared qubits}

\author{Konrad Banaszek}

\address{Center for Quantum Information, University of Rochester,
Rochester, New York, 14627\\
and Centre for Quantum Computation, Clarendon Laboratory, University
of Oxford, Parks Road, Oxford OX1 3PU, United Kingdom\thanks{Present address}}

\author{Igor Devetak}

\address{Department of Electrical and Computer Engineering, Cornell University,
Ithaca, New York, 14850}

\date{\today}

\maketitle

\draft

\begin{abstract}
We calculate the trade-off between the quality of estimating the quantum
state of an ensemble of identically prepared qubits and the minimum
level of disturbance that has to be introduced by this procedure in
quantum mechanics. The trade-off is quantified using two mean fidelities:
the operation fidelity which characterizes the average resemblance of
the final qubit state to the initial one, and the estimation fidelity
describing the quality of the obtained estimate. We analyze properties of
quantum operations saturating the achievability bound for the operation
fidelity versus the estimation fidelity, which allows us to reduce
substantially the complexity of the problem of finding the trade-off
curve. The reduced optimization problem has the form of an eigenvalue
problem for a set of tridiagonal matrices, and it can be easily solved
using standard numerical tools.
\end{abstract}


\section{Introduction}

It is a well-known fact that given a single copy of a quantum system
it is in general impossible to determine its quantum state exactly 
\cite{SingleCopy}.  This principle is closely related to the no-cloning
theorem \cite{WootZureNAT82}, which prevents one from producing
multiple faithful copies from a single unknown state. Of course, the
situation becomes different when one is given an ensemble of identically
prepared quantum systems. With the increasing size of such an ensemble
one can extract more and more precise information on its preparation
\cite{MassPopePRL95}.

An important effect which usually accompanies an attempt to determine
the unknown preparation of a quantum system is the disturbance
of its original state. The state disturbance is a penalty
for gaining classical information from a quantum system. This
fundamental feature of quantum mechanics has been discussed
from many different points of view, depending on the particular
physical scenario considered \cite{TradeOff,FuchJacoPRA01}. A
recent paper \cite{BanaPRL01} presented the description of the
trade-off between the information gain and the quantum state
disturbance motivated by the problem of quantum state estimation
\cite{MassPopePRL95,DerkBuzePRL98,LatoPascPRL98,VidaLatoPRA99,AcinLatoPRA00}.
In this approach, the classical outcome gained from an operation on the
quantum system is converted into a guess what the original state was.
According to the general quantum mechanical rule linking the information
gain with the state disturbance, the better guess can be made on
average, the less the final state of the system should resemble the
initial one.  Natural parameters for quantifying the trade-off in this
context are mean fidelities, defined using the scalar product between
the relevant state vectors, and averaged over many realizations of the
scenario. The first quantity of interest is the operation fidelity $F$,
which parameterizes the average resemblance of the state of the system
after the operation to the original one. The second quantity, known as
the estimation fidelity $G$, tells us how good estimate one can provide
on the basis of the classical outcome of a given operation. In the plane
of the fidelities $F$ and $G$, quantum mechanics imposes a bound in the
form of a trade-off curve limiting the maximum values of $F$ and $G$
that can be achieved simultaneously for any quantum operation. For
a single copy of a $d$-level system, the trade-off curve between $F$
and $G$ can be described in simple analytical terms \cite{BanaPRL01}.

In this paper we discuss a more general case of the fidelity trade-off,
when one is given a finite ensemble of identically prepared systems.
We shall assume that these systems are qubits prepared in an arbitrary,
randomly selected, pure state. Of course, the penalty --- i.e.\ the
state disturbance --- for gaining information is expected to be smaller
for an ensemble compared to the single system case. In the limit of an
infinite number of copies, we can determine the quantum state exactly
and reset the state of all the qubits according to this precisely known
information. Then in principle no disturbance needs to occur. Our interest
here will be focused on the intermediate case between the single-copy
operations and nearly perfect estimation of large ensembles. We shall
discuss the trade-off curve between the operation fidelity and the
estimation fidelity in the most general case when the ensemble consists
of a finite number $N$ of qubits.

The two fidelities used in our work have fundamentally different practical
meanings. Informally speaking, the operation fidelity $F$ deals with
the intrinsic quantum information remaining in the state of a quantum
system. In contrast, the estimation fidelity $G$ describes the classical
information gained from the measurement. This classical information
allows us, for example, to generate arbitrarily many copies of the initial
state with the same fidelity $G$. Consequently, for finite ensembles the
fidelities $F$ and $G$ take values from different ranges.  For example
$F$ can easily be equal to one for any size of the ensemble (which
simply means that nothing is done to the qubits) but in contrast $G=1$
means that we are able to generate arbitrarily many perfect copies using
the result of the estimation. Previous work on quantum state estimation
\cite{MassPopePRL95} demonstrated that the maximum attainable estimation
fidelity for an ensemble consisting of $N$ qubits is $G=(N+1)/(N+2)$. As
we justify later, this is also the value for the operation fidelity in
the limit of optimal quantum state estimation. Of interest here is what
happens below this value for $G$ and above this value for $F$, as this
describes the region where we try to convert part of the the quantum
information contained in the initial state into a classical guess.

Finding the trade-off curve for an ensemble of $N$ qubits presents in
principle a rather complicated problem. A general quantum operation can map
the initial state of the qubits onto the full Hilbert space with the
dimensionality $2^N$, which grows exponentially with the size of the
ensemble. Furthermore, the classical operation outcomes used for the
estimation can a priori assume values from an arbitrarily large set.
In this paper we demonstrate in several steps that it is possible to
reduce the general problem of finding the trade-off curve to ${\cal
O}(\sqrt{N})$ independent constrained optimization problems, each
involving ${\cal O}(N)$ real variables. The reduced optimization
problems have quadratic form, and they can be solved numerically by
finding the eigenvectors of certain tridiagonal matrices. This is a
substantial reduction of the complexity of the problem compared to its
original formulation, which allows us to deal numerically with much
larger ensembles. Furthermore, numerical evidence strongly suggests
that in general just one from ${\cal O}(\sqrt{N})$ optimization problems
gives the full trade-off curve, but a strict mathematical proof of this
conjecture is lacking. Most importantly, we achieve the reduction of
the complexity without imposing any restrictions on the generality of
quantum operations considered, and the calculated trade-off curves are
both tight and universal.

\begin{figure}

\vspace*{\fill}

\begin{center}
\epsfig{file=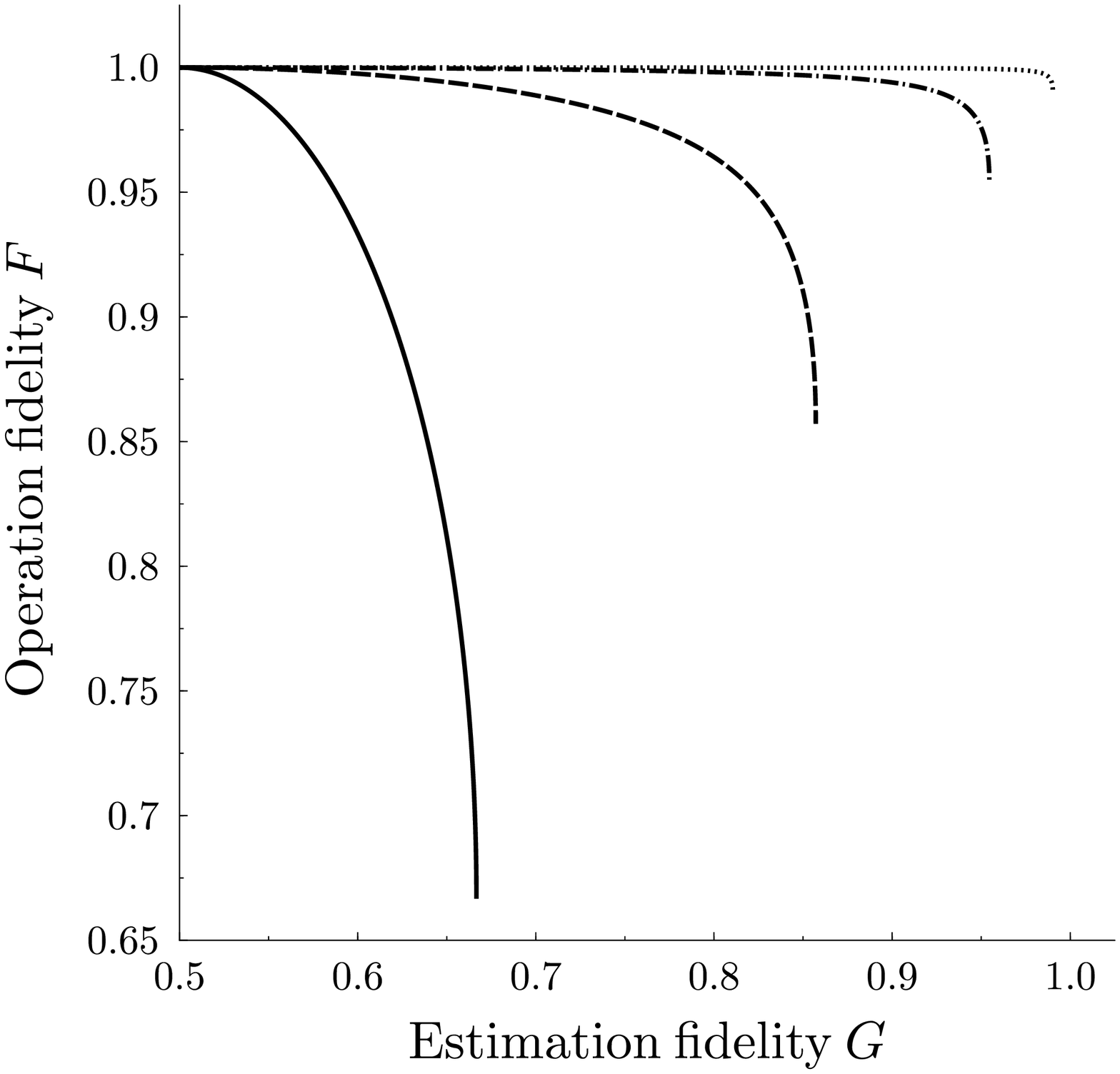,width=4in}
\epsfig{file=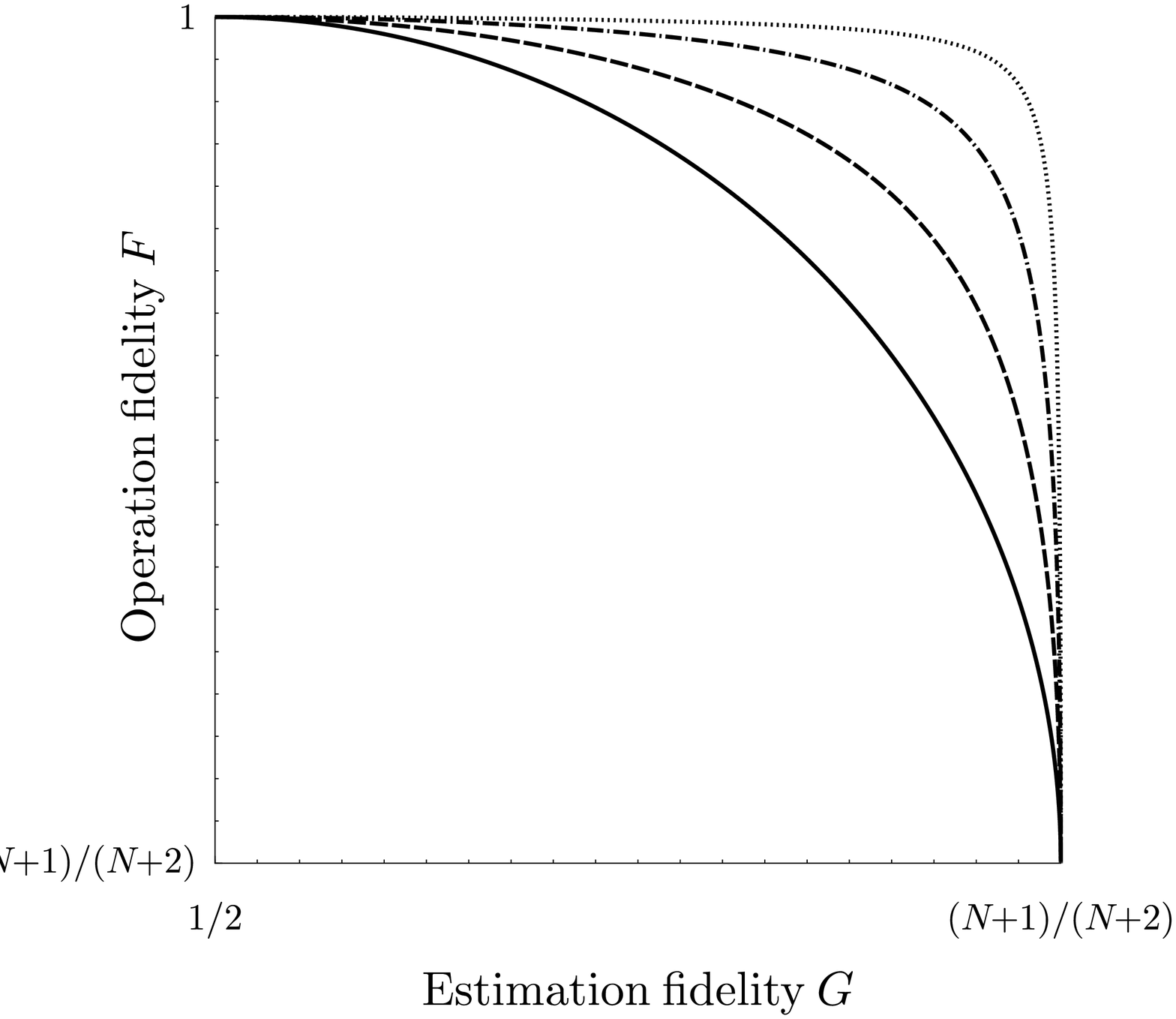,width=4in}
\end{center}

\vspace*{\fill}

\caption{The trade-off curves between the operation fidelity $F$
and the operation fidelity $G$, depicted using absolute (top)
and relative (bottom) scaling of the axes,
 for several values of the ensemble
size: $N=1$ (solid), $N=5$ (dashed), $N=20$ (dotted-dashed), and
$N=100$ (dotted).}
\label{Fig:Results}
\end{figure}

The results of our paper are summarized in Fig.~\ref{Fig:Results},
where we depict the trade-off curves calculated using our approach for
several exemplary values of $N$. All the curves have a common extreme
point attained for $F=1$ and $G=1/2$.  This point is reached when the
ensemble is simply left intact, and the corresponding value for the
estimation fidelity $G=1/2$ describes making a completely random guess
about the state of the qubits. The other extreme point for each trade-off
curve, corresponding to optimal quantum state estimation, is given by
$F=G=(N+1)/(N+2)$. In this limit, the fidelity of the qubits remaining
after the operation is exactly the same as the fidelity of our guess. One
can give a simple intuitive argument that this should be the case: $F$
is always an upper bound on $G$, since one can always set the state of
the qubits equal to the guess state. It is plausible that this bound
can be achieved in the limit where we only care about maximizing $G$.
Between the two extreme points, the depicted trade-off curves illustrate
how with increasing $N$ the extraction of classical information has less
of an effect on the state of the qubits after the operation.

The problem considered in this paper can be viewed as a special case
of quantum cloning, i.e.\ generating a larger number of imperfect
copies from a given ensemble \cite{BuzeHillPRA96}. As we noted, given
a classical estimate of the quantum state, we can use it to generate
an infinite number of qubits with the same fidelity as the estimate
\cite{BrusEkerPRL98}. Thus, the trade-off curves presented in this
paper describe the optimal performance of an asymmetric quantum copying
machine which given $N$ identical pure qubits produces $N$ clones with
the fidelity $F$, and in addition arbitrarily many clones with the
fidelity $G$.

The paper is organized as follows. First, in Sec.~\ref{Sec:Formulation},
we formulate the problem of the fidelity trade-off in quantitative terms.
In Sec.~\ref{Sec:Fidelities} we simplify the formulae for the fidelities
using the angular momentum representation of the rotation group. This
provides explicit expressions for the fidelities that are suitable for
further calculations. In Sec.~\ref{Sec:Decomposition} we argue that
in order to find the trade-off curve it suffices to consider a single
quantum operation element, thus substantially reducing the complexity
of the problem. We further demonstrate in Sec.~\ref{Sec:FullySym} that
it is sufficient to consider operations which map the initial state of
the qubits only onto the fully symmetric subspace. With these results in
hand, we define in Sec.~\ref{Sec:Optimization} the reduced optimization
problem which yields the actual trade-off curve. The numerical solution
to this problem is discussed in Sec.~\ref{Sec:Numerical}. Next, we show
in Sec.~\ref{Sec:Attainability} that the calculated trade-off curves
are achievable, by constructing explicit quantum operations attaining
the derived bound. Sec.~\ref{Sec:Discussion} concludes the paper.

\section{Formulation of the problem}
\label{Sec:Formulation}

We begin with an ensemble of $N$ qubits all prepared in the same
pure state $|\Omega\rangle$. We shall use the following notation:
\begin{equation}
|\Omega\rangle = \hat{U}(\Omega) | \uparrow \rangle,
\end{equation}
i.e.\ the state $|\Omega\rangle$ is represented as a result of a unitary
operation $\hat{U}(\Omega)$ on a reference state $|\uparrow\rangle$,
which we will take for concreteness to be the spin up state along the $z$
axis, $\hat{\sigma}^{z} | \uparrow \rangle = | \uparrow \rangle$.
The group of the unitary transformations $\hat{U}(\Omega)$ can
be conveniently parameterized using the two-dimensional irreducible
representation of the rotation group. Thus $\Omega$ can be considered
as an abbreviation for the triplet of the Euler angles $(\phi,
\theta, \zeta)$, and
\begin{equation}
\hat{U}(\Omega) = \exp(-i\phi \hat{\sigma}^z/2)
\exp(-i\theta \hat{\sigma}^y/2) \exp(-i\zeta \hat{\sigma}^z/2).
\end{equation}
The third Euler angle $\zeta$ introduces a trivial overall phase factor in
the definition of the states $|\Omega\rangle$ and in principle it could
be set to zero. However we will keep it as an independent variable in order
to adhere strictly to standard angular momentum algebra notation which we will
use later. The canonical volume element in the group of unitary transformations
$\hat{U}(\Omega)$ is given by
\begin{equation}
d\Omega = \frac{1}{8\pi^2} \sin\theta \, d\theta \, d\phi \, d\zeta.
\end{equation}
We assume that the initial state $|\Omega\rangle$ of the ensemble
of $N$ qubits is randomly selected according to the probability distribution
given by this measure.

Initially, the composite state of the ensemble of the qubits is described
by a tensor product $|\Omega\rangle \langle \Omega |^{\otimes N}$.
We assume that the qubits are submitted to an action of a certain
quantum operation, which can in general act collectively on the whole
ensemble. Such an operation is described by a set of operators $\{
\hat{A}_{rs} \}$ acting in the $2^N$-dimensional tensor product Hilbert
space of all the qubits \cite{Ozawa}.
The classical outcome of the operation is given
by the index $r$, and it is correlated with the final quantum state of
the qubits. The probability $p_r(\Omega) $ of obtaining the result $r$
is given by:
\begin{equation}
\label{Eq:pr}
p_r (\Omega) 
= \sum_{s} \text{Tr} ( \hat{A}_{rs}^\dagger \hat{A}_{rs}
|\Omega\rangle \langle \Omega |^{\otimes N} ).
\end{equation}
The conditional
transformation of the ensemble corresponding to the outcome $r$
is described by the formula:
\begin{equation}
\label{Eq:RhoOut}
\hat{\varrho}^{\text{out}}_r (\Omega) =
\frac{1}{p_r(\Omega)} \sum_{s} \hat{A}_{rs}
|\Omega\rangle \langle \Omega |^{\otimes N}
\hat{A}_{rs}^{\dagger}.
\end{equation}
The summation over the index $s$ maps in general pure states onto
mixed ones, and it can be viewed as responsible for introducing excess
stochastic fluctuations \cite{BarnumPhD}.  In general, the index $s$
can assume values from a different set for each $r$. For the operation
to be trace-preserving, the set of the operators $\{ \hat{A}_{rs} \}$
must satisfy the completeness relation of the form:
\begin{equation}
\label{Eq:Completeness}
\sum_{rs} \hat{A}_{rs}^{\dagger} \hat{A}_{rs} = \hat{\openone}.
\end{equation}

With the above notation for quantum operations, we can now define
explicitly the two quantities central to this paper: the operation
fidelity $F$ and the estimation fidelity $G$. The operation fidelity
$F$ quantifies the average resemblance of the state after the operation
to the original one. Let us consider the reduced single-qubit density
matrix after the operation, averaged over all $N$ qubits:
\begin{equation}
\hat{\varrho}^{\text{red}}_{r} (\Omega)
=
\frac{1}{N}
[
\text{Tr}_{2\mbox{-}N} \hat{\varrho}^{\text{out}}_{r} (\Omega)
+
\text{Tr}_{1,3\mbox{-}N} \hat{\varrho}^{\text{out}}_{r} (\Omega)
+ \ldots +
\text{Tr}_{1\mbox{-}(N-1)} \hat{\varrho}^{\text{out}}_{r} (\Omega)
],
\end{equation}
where the subscript of the trace symbol $\text{Tr}$ labels the range
of qubits over which the trace operation is performed. The expectation
value of the above expression on the initial state $\langle \Omega |
\hat{\varrho}^{\text{red}}_{r} (\Omega) | \Omega \rangle$ tells us how
much on average the state of a single qubit after the operation resembles
its initial value. This quantity, summed over the possible outcomes $r$
of the operation with the corresponding weights $p_r$ and averaged over
the initial state of the qubits, yields the mean operation fidelity for an
ensemble of identically prepared qubits:
\begin{equation}
\label{Eq:Fdef}
F = \int d\Omega \sum_{r} p_r (\Omega)
\langle \Omega | \hat{\varrho}^{\text{red}}_{r} (\Omega)
| \Omega \rangle.
\end{equation}

The second quantity of interest is the estimation fidelity $G$. Given the
classical outcome $r$ of the operation, we can make a guess $|\Omega_r
\rangle$ what the original state of the qubits was. A natural way
to quantify the quality of the guess is to take the squared absolute
value of the scalar product between the guess and the original state,
equal to $|\langle \Omega_r | \Omega \rangle|^2$.
The estimation fidelity is obtained by averaging this expression
over the sets of possible operation outcomes $r$ and the input states
$|\Omega\rangle$:
\begin{equation}
G = \int d\Omega \sum_{r} p_r (\Omega)
|\langle \Omega_r | \Omega \rangle|^2.
\end{equation}
The estimation fidelity depends not only on the quantum operation
$\{\hat{A}_{rs}\}$ itself, but also on the estimation rule used to make
the guess, described by the mapping $r \mapsto | \Omega_r \rangle$.
We will demonstrate in the following how to define this mapping in
a way which optimizes the estimation fidelity for a given arbitrary
quantum operation.

Our goal is now to find the inequality which bounds the fidelities $F$ and
$G$, assuming the most general form of the operation $\{ \hat{A}_{rs}
\}$. The only requirement that has to be satisfied by the set of the
operators $\hat{A}_{rs}$ is the completeness condition described by
Eq.~(\ref{Eq:Completeness}).

\section{Fidelities}
\label{Sec:Fidelities}

In this section, we will simplify the expression for the fidelities $F$
and $G$ to the form which makes them more manageable in the optimization
procedure.  Our first step will be explicit calculation of the integrals
over the space of pure states $|\Omega\rangle$, which can be done with
the help of tools developed in the theory of representations of the
rotation group.  Throughout this section, we shall follow strictly the
notation of Ref.~\cite{Brink} for the angular momentum algebra and the
elements of rotation matrices.

To simplify subsequent expressions, we begin with a general
observation that the index $s$ appearing in Eqs.~(\ref{Eq:pr}) and
(\ref{Eq:RhoOut}) could in principle also be known classically
after the operation. Summation over the index $s$ in Eqs.~(\ref{Eq:pr})
and Eq.~(\ref{Eq:RhoOut}) means that the operation is imperfect, and
it averages statistically different output states. This results in the loss
of a fraction of the information extracted from the initial quantum
state. As we are interested in the optimal operations saturating the
quantum-mechanical bound on the fidelities, we can assume with no loss
of generality that both $r$ and $s$ are known. In such a case, we can
use a new single index to label both $r$ and $s$. For this reason,
we will assume in the following that the index $s$ is trivial, i.e.\ it
assumes only a single value for each $r$, and that consequently it can be
dropped from further notation. Thus we restrict our attention to quantum
operations which are known in the literature as ideal \cite{NielCavePRA97}
or efficient \cite{FuchJacoPRA01}.

In the following calculations, it will be convenient to use the
decomposition of the complete Hilbert space of $N$ qubits into subspaces
with the fixed value of the total angular momentum operator.
This decomposition has the form \cite{CiraEkerPRL99}:
\begin{equation}
\label{Eq:HDecomposition}
\bigotimes_{i=1}^{N} {\cal H}
=
\bigoplus_{\alpha = 1}^{2j \choose {\lfloor j \rfloor}}
{\cal H}_{j_\alpha},
\end{equation}
where $j=N/2$ is the largest value of the angular momentum appearing in
the decomposition, the $j_\alpha$'s assume values in the set $\{j, j-1 ,
\ldots, j - \lfloor j \rfloor \}$ and are arranged in a non-ascending
order. Here ${\cal H}_{j'}$ denotes the subspace corresponding to the
total angular momentum value $j'$. We note that $j_1 = j$ and $j_\alpha <
j$ for $\alpha > 1$, as there is only one representation with $j=N/2$
corresponding to the fully symmetric subspace. For completeness, in
Appendix~\ref{Appendix:Multiplicities} we present a simple derivation
of the multiplicities of the angular momentum representations appearing
in Eq.~(\ref{Eq:HDecomposition}).  The action of a tensor product of
the unitary operators $\hat{U}^{\otimes N} (\Omega)$ in each of the
subspaces ${\cal H}_{j'}$ is given by the corresponding representation
of the rotation group.  In order to simplify the notation, we will
henceforth use the same symbol $\hat{U}(\Omega)$ to denote the action
of $\hat{U}^{\otimes N} (\Omega)$  on the whole ensembles of qubits,
as for a single qubit.

\subsection{Operation fidelity}
\label{SubSec:OperationFidelity}

Let us start with the operation fidelity $F$. The matrix
element $\langle\Omega | \hat{\varrho}_r^{\text{red}} (\Omega) | \Omega
\rangle$ appearing in the definition of $F$ in Eq.~(\ref{Eq:Fdef})
is equivalently given by the expectation value over the density
matrix $\hat{\varrho}_r^{\text{out}}(\Omega)$ defined in
Eq.~(\ref{Eq:RhoOut}) of the following operator:
\begin{equation}
\hat{P}(\Omega) = 
\frac{1}{N}
(|\Omega\rangle\langle\Omega| \otimes \hat{\openone}
\otimes \cdots \otimes \hat{\openone}
+
\hat{\openone} \otimes |\Omega\rangle\langle\Omega| 
\otimes \cdots \otimes \hat{\openone}
+
\hat{\openone} \otimes \hat{\openone}
\otimes \cdots \otimes |\Omega\rangle\langle\Omega|
)
\end{equation}
Using the above definition,
the mean estimation fidelity can be written compactly as:
\begin{equation}
F = \int d\Omega \sum_{r} \mbox{Tr} [
\hat{P}(\Omega) \hat{A}_r | \Omega \rangle
\langle \Omega | ^{\otimes N}
 \hat{A}_r^\dagger ].
\end{equation}
We will now express $\hat{P}(\Omega)$ in terms
of the angular momentum operators.
Expressing the projection $|\Omega\rangle\langle
\Omega |$ as a combination of the Pauli matrices:
\begin{equation}
|\Omega\rangle\langle
\Omega |
=
\frac{1}{2}
(
\hat{\openone} + \sin \theta
\cos\phi \, \hat{\sigma}^x + \sin\theta \sin\phi
\, \hat{\sigma}^y + \cos\theta \, \hat{\sigma}^z)
\end{equation}
we can write the operator $\hat{P}(\Omega)$ in terms of the total
angular momentum operators for the composite system of $N$ qubits:
\begin{eqnarray}
\hat{P}(\Omega) 
& = & \frac{1}{2N} \sum_{i=1}^{N} (\hat{\openone}
+ \sin\theta \cos\phi \, \hat{\sigma}^x_{i} + \sin\theta \sin\phi
\, \hat{\sigma}^y_{i} + \cos\theta \, \hat{\sigma}^z_{i})
\nonumber \\
& = &
\frac{1}{2} \hat{\openone} +
\frac{1}{N} ( \sin\theta \cos\phi \hat{J}^x +
\sin\theta \sin\phi \hat{J}^y
+ \cos\theta \hat{J}^z ),
\end{eqnarray}
where the index $i$ enumerates the qubits. In the following, it will 
be convenient to switch to the pair of the
angular momentum raising and lowering operators
$\hat{J}^{\pm} = \hat{J}^x \pm i \hat{J}^y$, using which we have:
\begin{equation}
\hat{P}(\Omega) 
 =  
\frac{1}{2} \hat{\openone} +
\frac{1}{N} \left( \frac{1}{2} e^{i\phi} \sin\theta \hat{J}^- +
\frac{1}{2} e^{-i\phi} \sin\theta \hat{J}^+
+ \cos\theta \hat{J}^z \right) 
\end{equation}
With this expression for the operator $\hat{P}(\Omega)$, we can
write the mean fidelity $F$ as:
\begin{equation}
\label{Eq:FwithK}
F = \frac{1}{2} + \frac{1}{N} \sum_{r} \mbox{Tr}
\left(
\frac{1}{2} \hat{J}^{-} \hat{A}_r \hat{K}_{-1} \hat{A}_r^\dagger
+ \frac{1}{2} \hat{J}^{+} \hat{A}_r \hat{K}_{1} \hat{A}_r^\dagger
+ \hat{J}^{z} \hat{A}_r \hat{K}_{0} \hat{A}_{r}^\dagger
\right),
\end{equation}
where all the terms involving the Euler angles $\Omega$ have been
collected to three integrals
$\hat{K}_\tau$, $\tau=-1,0,1$, defined as
\begin{equation}
\label{Eq:Kdef}
\hat{K}_{\tau}
=
\int d\Omega \,
k_\tau (\Omega)
| \Omega \rangle \langle \Omega | ^{\otimes N} 
\end{equation}
with the functions $k_\tau(\Omega)$ given by:
\begin{eqnarray}
k_{-1}(\Omega) & = &
e^{i\phi}\sin\theta
\nonumber \\
k_{0}(\Omega) & = &
\cos\theta
\nonumber \\
\label{Eq:kdef}
k_{1}(\Omega) & = &
e^{-i\phi}\sin\theta.
\end{eqnarray}
Explicit calculation of the integrals $\hat{K}_\tau$, performed in
Appendix~\ref{Appendix:K}, yields the following expressions:
\begin{eqnarray}
\hat{K}_{\pm 1} & = & \frac{1}{(j+1)(2j+1)} \hat{J}^{\mp}_{j}
\nonumber \\
\hat{K}_{0} & = & \frac{1}{(j+1)(2j+1)} \hat{J}^z_{j}
\end{eqnarray}
where $j=N/2$ and $\hat{J}^{\mp}_{j}, \hat{J}^z_{j}$ denote the angular
momentum operators restricted to the completely symmetric subspace
of the $N$ qubits defined by the angular momentum $j=N/2$. The operators
$\hat{K}_{\tau}$ vanish outside this space, as they are defined as
integrals of totally symmetric projection operators $|\Omega \rangle
\langle \Omega |^{\otimes N}$.

Inserting the explicit form of the operators $\hat{K}_{\tau}$ into
Eq.~(\ref{Eq:FwithK}) yields:
\begin{equation}
F = \frac{1}{2} + \frac{1}{2j(j+1)(2j+1)}
\sum_{r}
\mbox{Tr}
\left(
\frac{1}{2}
\hat{J}^{-} \hat{A}_r \hat{J}^{+}_{j} \hat{A}_r^{\dagger}
+ \frac{1}{2}
\hat{J}^{+} \hat{A}_r \hat{J}^{-}_{j} \hat{A}_r^{\dagger}
+ \hat{J}^{z} \hat{A}_{r} \hat{J}^{z}_{j} \hat{A}_{r}^{\dagger}
\right)
\end{equation}
The operation fidelity can be equivalently expressed in terms of
the Hermitian operators $\hat{J}^x$, $\hat{J}^{y}$, and $\hat{J}^{z}$ as:
\begin{equation}
F = \frac{1}{2} + \frac{1}{2j(j+1)(2j+1)}
\sum_r
\mbox{Tr}(
\hat{J}^x \hat{A}_r \hat{J}^{x}_{j} \hat{A}_r^\dagger
+
\hat{J}^y \hat{A}_r \hat{J}^{y}_{j} \hat{A}_r^\dagger
+
\hat{J}^z \hat{A}_r \hat{J}^{z}_{j} \hat{A}_r^\dagger
)
\end{equation}
We note that this expression is completely symmetric with respect to
the three Cartesian components of the angular momentum. Furthermore,
it can be easily checked
that each of the traces appearing in the sum
over $r$ in the above formula is invariant with respect to the transformation
of the operator $\hat{A}_r$ according to:
\begin{equation}
\hat{A}_r \rightarrow \hat{U}(\Omega) \hat{A}_r \hat{U}^\dagger (\Omega),
\end{equation}
where $\hat{U}(\Omega)$ is an arbitrary rotation matrix.

\subsection{Estimation fidelity}

We will now evaluate the integral over the space of pure states in the
expression for the estimation fidelity $G$, given by:
\begin{equation}
G = \int d\Omega \sum_{r} 
\mbox{Tr} (\hat{A}_r^\dagger \hat{A}_r 
| \Omega \rangle \langle \Omega |^{\otimes N} 
)
| \langle \Omega_r | \Omega \rangle |^2
\end{equation}
As we discuss in Appendix~\ref{Appendix:K}, the projection on the
product state $| \Omega \rangle \langle \Omega |^{\otimes N}$ can
be represented as:
\begin{equation}
| \Omega \rangle \langle \Omega |^{\otimes N}
= \hat{U} (\Omega) | j;j \rangle \langle j ; j | \hat{U}^\dagger
(\Omega),
\end{equation}
where $j=N/2$
and $|j;j\rangle$ belongs to the subspace with the total
angular momentum $j$ and is the eigenvector of $\hat{J}^z$ corresponding
to the eigenvalue $j$.
Thus we have:
\begin{equation}
G = \int d\Omega \sum_{r} 
| \langle \uparrow | \hat{U}^\dagger(\Omega_r) 
\hat{U}(\Omega)
| \uparrow \rangle |^2
\mbox{Tr} [ \hat{A}_r^\dagger \hat{A}_r 
\hat{U}(\Omega)
| j; j \rangle \langle j ; j |
\hat{U}^{\dagger}(\Omega)
]
\end{equation}
We can now swap the order of the integration over $\Omega$ and the 
summation over $r$, and change for each $r$ the integration variables
from  $\Omega$ to $\Omega'$ such that:
\begin{equation}
\hat{U}(\Omega') =
\hat{U}^\dagger(\Omega_r) 
\hat{U}(\Omega)
\end{equation}
Using this parameterization, we have $\langle \uparrow |
\hat{U}^\dagger(\Omega_r) \hat{U}(\Omega) | \uparrow \rangle |^2
= (1+ \cos\theta')/2$ and
\begin{equation}
G = \frac{1}{2} \sum_{r} \int d\Omega' \, (1 + \cos\theta')
\mbox{Tr} [ \hat{A}_r^\dagger
\hat{A}_{r} \hat{U} (\Omega_r) \hat{U} (\Omega') | j;j \rangle
\langle j ; j | \hat{U}^{\dagger} (\Omega') 
\hat{U}^\dagger(\Omega_r) ]
\end{equation}
This expression can be decomposed into two parts according to
the two terms in the factor $1+ \cos\theta'$. The first part
involves the integral
\begin{equation}
\int d\Omega' \, \hat{U}(\Omega') | j;j \rangle
\langle j ; j | \hat{U}^{\dagger} (\Omega')
= \frac{1}{2j+1} \hat{\openone}_{j}
\end{equation}
which is proportional to the identity operator $\hat{\openone}_{j}$
truncated to the completely symmetric subspace. Consequently, the summation
over $r$ for this term can be easily performed which yields
the constant value $1/2$. The second
part can be expressed with the help of the operator $\hat{K}_{0}$ defined
in Eq.~(\ref{Eq:Kdef}), which gives:
\begin{equation}
G = \frac{1}{2} \sum_{r} \frac{\text{Tr}(
\hat{A}^\dagger_r \hat{A}_r \hat{\openone}_j )}{2j+1}
+ \frac{1}{2}
\sum_{r} \text{Tr}
[
\hat{A}_r^\dagger
\hat{A}_{r} \hat{U} (\Omega_r) \hat{K}_{0}
\hat{U}^\dagger(\Omega_r) 
]
\end{equation}
Inserting the explicit form of $\hat{K}_0$ yields:
\begin{equation}
\label{Eq:G=sumrTr}
G = \frac{1}{2} + \frac{1}{2(j+1)(2j+1)} \sum_{r} \mbox{Tr}
[ \hat{A}_r^\dagger \hat{A}_{r}
\hat{U}(\Omega_r) \hat{J}^z_{j} 
\hat{U}^{\dagger}(\Omega_r) ]
\end{equation}
This expression for the estimation fidelity allows one to derive easily
the optimal estimation strategy, i.e.\ the mapping from the set of outcomes
$r$ to guesses $\Omega_r$ for a given quantum operation $\{\hat{A}_r\}$.
In order to derive this strategy we first note that
\begin{equation}
\hat{U}(\Omega_r) \hat{J}^z_{j}
\hat{U}^{\dagger}(\Omega_r)
= \sin\theta_r \cos\phi_r \hat{J}^{x}_{j}
+ \sin\theta_r \sin\phi_r \hat{J}^{y}_{j}
+ \cos \theta_r \hat{J}^{z}_{j}.
\end{equation}
This allows us to write each of the traces in the sum over $r$ in the form
of a scalar product ${\bf A}_r^{T} {\bf \Omega}_{r}$ between two
three-dimensional real vectors ${\bf A}_r$ and ${\bf \Omega}_{r}$
defined as:
\begin{equation}
\label{Eq:vecAdef}
{\bf A}_r = \left(
\begin{array}{c}
\text{Tr} ( \hat{A}_r^\dagger \hat{A}_{r} \hat{J}^{x}_{j} ) \\
\text{Tr} ( \hat{A}_r^\dagger \hat{A}_{r} \hat{J}^{y}_{j} ) \\
\text{Tr} ( \hat{A}_r^\dagger \hat{A}_{r} \hat{J}^{z}_{j} )
\end{array}
\right)
\end{equation}
and
\begin{equation}
{\bf\Omega}_{r}= \left(
\begin{array}{c}
\sin\theta_{r} \cos\phi_{r} \\
\sin\theta_{r} \sin\phi_{r} \\
\cos \theta_{r}
\end{array}
\right)
\end{equation}
It is seen that ${\bf\Omega}_{r}$ is a unit vector pointing in the
direction defined by the Euler angles $\Omega_r$.
Obviously, the scalar product ${\bf A}_r^T {\bf\Omega}_{r}$ is maximized
when the two vectors are parallel. Thus, for a given $r$ the corresponding
Euler angles $\phi_r$ and $\theta_r$ should be chosen such that:
\begin{equation}
{\bf \Omega}_{r} = \frac{{\bf A}_r}{\sqrt{{\bf A}_r^T {\bf A}_r}}.
\end{equation}
Of course, the value of the third Euler angle $\zeta_r$ can be arbitrary,
as it introduces only an irrelevant overall phase factor. If the vector
${\bf A}_{r}$ is zero, then the guess has to be made completely randomly.
Thus, the maximum
value of the trace for a specific $r$ in Eq.~(\ref{Eq:G=sumrTr}) is given by
$\sqrt{{\bf A}_r^T {\bf A}_r}$, and the maximum value of the estimation
fidelity attainable for a given operation $\{\hat{A}_r\}$ is defined by:
\begin{equation}
G = \frac{1}{2} + \frac{1}{2(j+1)(2j+1)} \sum_{r} 
\sqrt{
[ \text{Tr} ( \hat{A}_r^\dagger \hat{A}_{r} \hat{J}^{x}_{j} ) ]^2 +
[ \text{Tr} ( \hat{A}_r^\dagger \hat{A}_{r} \hat{J}^{y}_{j} ) ]^2 +
[ \text{Tr} ( \hat{A}_r^\dagger \hat{A}_{r} \hat{J}^{z}_{j} ) ]^2
}.
\end{equation}
As in the case of the operation fidelity, this expression is invariant with
respect to an arbitrary transformation of the operators $\hat{A}_r$ according
to the rotation group.

\section{Decomposition}
\label{Sec:Decomposition}

Let us now summarize the constrained optimization problem describing
the trade-off between the operation fidelity $F$ and the estimation fidelity
$G$. These two quantities can be written as:
\begin{equation}
F = \frac{1}{2} + \frac{1}{2j(j+1)} f
\end{equation}
and
\begin{equation}
G = \frac{1}{2} + \frac{1}{2(j+1)} g
\end{equation}
where the terms $f$ and $g$ depending explicitly on the operation
$\{ \hat{A}_r \}$ are given by:
\begin{equation}
\label{Eq:fdef}
f = \frac{1}{2j+1} \sum_{r}
\text{Tr} ( \hat{J}^{x} \hat{A}_r \hat{J}^{x}_{j} \hat{A}_{r}^{\dagger}
+
\hat{J}^{y} \hat{A}_r \hat{J}^{y}_{j} \hat{A}_{r}^{\dagger}
+
\hat{J}^{z} \hat{A}_r \hat{J}^{z}_{j} \hat{A}_{r}^{\dagger})
\end{equation}
and
\begin{equation}
\label{Eq:gdef}
g = \frac{1}{2j+1} \sum_{r} 
\sqrt{
[ \text{Tr} ( \hat{A}_r^\dagger \hat{A}_{r} \hat{J}^{x}_{j} ) ]^2 +
[ \text{Tr} ( \hat{A}_r^\dagger \hat{A}_{r} \hat{J}^{y}_{j} ) ]^2 +
[ \text{Tr} ( \hat{A}_r^\dagger \hat{A}_{r} \hat{J}^{z}_{j} ) ]^2
}
\end{equation}
It is worthwhile to look first at the extreme cases. The maximum value of
the operation fidelity itself is of course $F=1$. This corresponds
to $f=j(j+1)$.  This limit is achieved by the identity operation,
for which it is easy to check that $g=0$. Hence we can assume in the
following that $g$ is a nonnegative quantity. The other extreme case is
the optimization of the estimation fidelity alone, which has been studied
previously \cite{MassPopePRL95}.  According to these results, in the
limit of optimal quantum estimation we obtain $g=j$, which sets the
upper bound on the region of interest for $g$. We will demonstrate in
Sec.~\ref{Sec:Numerical} that the maximum value of $f$ attainable in
the case of optimal estimation is equal to $f=j^2$. Expressing this in
terms of the fidelities, we get that $F=G$. 

The operators $\hat{A}_r$ must satisfy in general the completeness condition
described in Eq.~(\ref{Eq:Completeness}). However, since the initial state
of the $N$ qubits lies in the completely symmetric subspace, we need to
consider the action of these operators only on the fully
symmetric subspace described by ${\cal H}_j$.
Consequently for the purpose of our discussion
we shall assume that $\hat{A}_r : {\cal H}_{j} \rightarrow
{\cal H}^{\otimes N}$,
and that the trace preserving condition takes the form:
\begin{equation}
\label{Eq:Trpresj}
\sum_{r} \hat{A}_r^\dagger \hat{A}_r = \hat{\openone}_j.
\end{equation}
Henceforth we replace the trace preserving condition by the milder requirement,
obtained by taking the trace of the above equation:
\begin{equation}
\label{Eq:WeakerCompleteness}
\text{Tr} \left( \sum_{r} \hat{A}_{r}^{\dagger} \hat{A}_{r} \right) = 2j+1.
\end{equation}
We will perform the optimization under this weakened constraint and
then show that the optimal $(f,g)$ curve may be attained by a set of
$\hat{A}_r$ that is actually trace preserving.

Weakening the completeness condition allows us to introduce an
important simplification in further calculations. As we noted in
Sec.~\ref{SubSec:OperationFidelity}, the expression for the operation
fidelity is invariant with respect to rotations performed on the
operators $\hat{A}_r$. This is also the case of the weaker trace
preserving condition described in Eq.~(\ref{Eq:WeakerCompleteness}).
Consequently, we can always modify each of the operators $\hat{A}_r$
by an operation of the form $ \hat{A}_r \rightarrow \hat{U}(\Omega)
\hat{A}_r \hat{U}^\dagger (\Omega)$ such that the vector defined in
Eq.~(\ref{Eq:vecAdef}) is aligned along the $z$ axis and moreover its
$z$ component is nonnegative. In explicit terms, we assume that
$\text{Tr} (\hat{A}_r^\dagger \hat{A} \hat{J}^{x}_{j}) =
\text{Tr} (\hat{A}_r^\dagger \hat{A} \hat{J}^{y}_{j})
= 0$, and $\text{Tr} (\hat{A}_r^\dagger \hat{A} \hat{J}^{z}_{j})
\ge 0$. This allows us to replace the square
root appearing in Eq.~(\ref{Eq:gdef}) by a much simpler expression
$\text{Tr}(\hat{A}_r^\dagger \hat{A}_r \hat{J}^{z}_{j})$. An 
important advantage of this step is that the latter expression
is quadratic in the matrix elements of the operators $\hat{A}_r$.

Our next step will be the representation of $f$ and $g$ as linear
combinations:
\begin{eqnarray}
f & = & \sum_{r} \lambda_r f(\hat{A}_{r}) \nonumber \\
g & = & \sum_{r} \lambda_r g(\hat{A}_{r})
\end{eqnarray}
where
\begin{mathletters}
\begin{eqnarray}
\label{Eq:fAdef}
f (\hat{A}) & = & \frac{1}{\text{Tr} (\hat{A}^\dagger \hat{A} )}
\text{Tr} ( \hat{J}^{x} \hat{A} \hat{J}^{x}_{j} \hat{A}^{\dagger}
+
\hat{J}^{y} \hat{A} \hat{J}^{y}_{j} \hat{A}^{\dagger}
+
\hat{J}^{z} \hat{A} \hat{J}^{z}_{j} \hat{A}^{\dagger})
\\
\label{Eq:gAdef}
g (\hat{A}) & = & \frac{1}{\text{Tr} (\hat{A}^\dagger \hat{A} )}
\text{Tr}(\hat{A}^\dagger \hat{A} \hat{J}^{z}_{j})
\end{eqnarray}
\end{mathletters}
and the nonnegative coefficients $\lambda_r$ are given by
\begin{equation}
\lambda_r = \frac{\text{Tr} (\hat{A}_r^\dagger \hat{A}_r )}{2j+1}.
\end{equation}
The values of $(f(\hat{A}_r), g(\hat{A}_{r}))$ are insensitive to
the rescaling of $\hat{A}_{r}$, so $\lambda_r$ are free variables
up to the constraint
\begin{equation}
\sum_{r} \lambda_r = 1
\end{equation}
resulting from the weaker trace preserving condition.  Hence each $(f,g)$
point is a {\em convex} combination of independent $(f(\hat{A}_r),
g(\hat{A}_{r}))$. As we are interested in the boundary of allowed
$(f,g)$ points, it suffices to examine the case when the $\lambda_r$
are all zero except for one value of $r$. This means that our problem
is solved by using only one operation element which we denote
by $\hat{A}$, obeying the constraint $\text{Tr} (\hat{A}^\dagger
\hat{A} ) = j + 1$. This observation has also been used in quantum
rate-distortion theory \cite{QRateDistortion}. After deriving the bound
for $(f,g)$ based on a single element $\hat{A}$, we will demonstrate in
Sec.~\ref{Sec:Attainability} that the operator $\hat{A}$ can be used in
a canonical way to construct a quantum operation satisfying the original
full completeness condition.

\section{Fully symmetric subspace}
\label{Sec:FullySym}

We will now show that the optimization problem can be simplified
even further: namely, that it is sufficient to consider operators
$\hat{A}$ which do not transfer the state of $N$ qubits beyond the fully
symmetric subspace.

According to our discussion of the structure of the Hilbert space of
the whole ensemble, the operator $\hat{A} : {\cal H}_j \rightarrow
{\cal H}^{\otimes N}$ may be
viewed as consisting of entries $\hat{A}_{\alpha}$,
each entry acting from $\hat{A}_{\alpha} : {\cal H}_j \rightarrow
{\cal H}_{j_\alpha}$. As the angular momentum operators are block
diagonal and they do not mix subspaces with different $\alpha$'s,
we can introduce the following decomposition of the quantities
$f$ and $g$:
\begin{equation}
f = \sum_{\alpha = 1}^{
{ 2j \choose {\lfloor j \rfloor }}
} \lambda_\alpha f_{j_\alpha}
( \hat{A}_{\alpha} )
\end{equation}
and
\begin{equation}
g = \sum_{\alpha = 1}^{
{ 2j \choose {\lfloor j \rfloor }}
} \lambda_\alpha g_{j_\alpha}
( \hat{A}_{\alpha} )
\end{equation}
where the positive coefficients $\lambda_\alpha$ are given by:
\begin{equation}
\lambda_\alpha = \text{Tr} (\hat{A}_{\alpha}^{\dagger} \hat{A}_{\alpha})
\end{equation}
and the functions $f_{j'} (\hat{B})$ and $g_{j'} (\hat{B}) $ are defined
as:
\begin{eqnarray}
f_{j'} (\hat{B}) & = & \frac{1}{\text{Tr}(\hat{B}^\dagger \hat{B})}
\text{Tr} (
\hat{J}^x_{j'} \hat{B} \hat{J}^{x}_{j} \hat{B}^{\dagger}
+
\hat{J}^y_{j'} \hat{B} \hat{J}^{y}_{j} \hat{B}^{\dagger}
+
\hat{J}^z_{j'} \hat{B} \hat{J}^{z}_{j} \hat{B}^{\dagger}
) \nonumber \\
g_{j'} (\hat{B}) & = &
\frac{1}{\text{Tr}(\hat{B}^\dagger \hat{B})}
\text{Tr}( \hat{B}^\dagger \hat{B} \hat{J}^{z}_{j})
\end{eqnarray}
Here $\hat{J}^{x}_{j'}$, $\hat{J}^{y}_{j'}$, and $\hat{J}^{z}_{j'}$
denote the angular momentum operators truncated to the subspace
${\cal H}_{j'}$.
As before, the values of $(f_{j_\alpha}(\hat{A}_{\alpha}), 
g_{j_\alpha} (\hat{A}_{\alpha}))$ are insensitive to the rescaling of
$\hat{A}_{\alpha}$ by a multiplicative factor, so 
consequently $\lambda_\alpha$ are free variables up to the constraint:
\begin{equation}
\sum_{\alpha=1}^{
{ 2j \choose {\lfloor j \rfloor }}
} \lambda_\alpha = 1.
\end{equation}
Hence again each $(f,g)$ point
is a {\em convex} combination of independent
$(f_{j_\alpha}(\hat{A}_{\alpha}),g_{j_\alpha} (\hat{A}_{\alpha}))$.
Consequently, it is sufficient to examine the much simpler case
when all the $\lambda_\alpha$ are all zero except for one value
of $\alpha$. This means that out problem reduces to finding
the upper boundaries of individual $(f_{j'}, g_{j'})$ regions,
where $ j' \in { j, j-1, \ldots, j - \lfloor j \rfloor}$.

Henceforth we drop the index $\alpha$ and consider a single operator
$\hat{A}_{j'}$ mapping the fully symmetric subspace ${\cal H}_j$ onto a
certain subspace ${\cal H}_{j'}$ with the total angular momentum equal to
$j'$. We can also assume with no loss of generality
that the operator $\hat{A}_{j'}$ is normalized
in such a way that:
\begin{eqnarray}
1 & = &
\text{Tr} ( \hat{A}_{j'}^\dagger \hat{A}_{j'} ) \nonumber \\
& = &
\sum_{m' = -j'}^{j'} \sum_{m = -j}^{j}
| \langle j' ; m' | \hat{A}_{j'} | j ; m \rangle | ^2 .
\end{eqnarray}
The explicit
expressions for the functions $f_{j'} (\hat{A}_{j'})$ and $g_{j'}
(\hat{A}_{j'})$ take then the following form:
\begin{eqnarray}
f_{j'} ( \hat{A}_{j'} ) & = & \sum_{m' =-j'}^{j'} \sum_{m = -j}^{j} m'm
| \langle j' ; m' | \hat{A}_{j'} | j ; m \rangle | ^2 \nonumber \\
& & + \sum_{m'=-j'}^{j'-1} \sum_{m=-j}^{j-1}
\sqrt{(j'+m'+1)(j'-m')(j+m+1)(j-m)} \nonumber \\
\label{Eq:fj'Aj'}
& & \times
\text{Re} [ \langle j'; m' | \hat{A}_{j'} | j ; m \rangle
(\langle j' ; m' +1 | \hat{A}_{j'} | j ; m+1 \rangle )^{\ast} ]
\end{eqnarray}
and
\begin{equation}
g_{j'} ( \hat{A}_{j'} ) = \sum_{m' = -j'}^{j'}
\sum_{m=-j}^{j} m 
| \langle j' ; m' | \hat{A}_{j'} | j ; m \rangle |^2 .
\end{equation}
As the next step to simplify the problem we note that the phases
of the matrix elements $\langle j'; m' | \hat{A}_{j'} | j ; m \rangle$
can be set to make all of them real and nonnegative. This maximizes
the second term in Eq.~(\ref{Eq:fj'Aj'}) while leaving unchanged
the expressions for $g_{j'}(\hat{A}_{j'})$ and $\text{Tr}
(\hat{A}_{j'}^\dagger \hat{A}_{j'})$.

We will now demonstrate that among all the curves bounding
the allowed regions of $(f_{j'}, g_{j'})$, the curve for $j'=j$
encompasses the largest region in the $(f,g)$ plane,
which includes all other bounds obtained for
$j' < j$. For this purpose, we will show that given an arbitrary
operator $\hat{A}_{j'} : {\cal H}_j \rightarrow {\cal H}_{j'}$
satisfying the condition $g_{j'} (\hat{A}_{j'}) \ge 0$,
it is possible to construct an operator $\hat{A}' : {\cal H}_{j}
\rightarrow {\cal H}_j$ mapping the fully symmetric space such that:
\begin{mathletters}
\begin{eqnarray}
\label{Eq:Condforf}
f_{j} (\hat{A}') & \ge & f_{j'} (\hat{A}_{j'}) \\
\label{Eq:Condforg}
g_{j} (\hat{A}') & = & g_{j'} (\hat{A}_{j'}) \\
\label{Eq:CondforTr}
\text{Tr} (\mbox{$\hat{A}'$}^{\dagger} \hat{A}') & = &
\text{Tr} (\hat{A}_{j'}^\dagger \hat{A}_{j'})
\end{eqnarray}
\end{mathletters}
Hence the operator $\hat{A}'$ will be always more optimal that
the original operator $\hat{A}_{j'}$.

The explicit construction of the operator $\hat{A}'$ is given by:
\begin{equation}
\langle j ; n | \hat{A}' | j ; m \rangle
= 
\left\{
\begin{array}{ll}
\langle j' ; n - j + j' | \hat{A}_{j'} | j ; m \rangle , &
\mbox{if $n \ge j - 2j'$}
\\
0 , & \mbox{if $n < j-2j'$}
\end{array}
\right.
\end{equation}
It is straightforward to check that the operator $\hat{A}'$ defined
above automatically satisfies conditions given by Eqs.~(\ref{Eq:Condforg})
and (\ref{Eq:CondforTr}). In order to prove that condition
(\ref{Eq:Condforf}) is also satisfied, 
let us express $f_j (\hat{A}')$ in terms of the matrix elements of
the operator $\langle j' ; m' | \hat{A}_{j'} | j ; m \rangle$:
\begin{eqnarray}
f_{j} (\hat{A}') & = &
\sum_{m' = -j'}^{j'} \sum_{m=-j}^{j}
(m'+j-j')m  | \langle j' ; m' | \hat{A}_{j'} | j ; m \rangle |^2 \nonumber \\
& & + \sum_{m'=-j'}^{j'-1} \sum_{m=-j}^{j-1}
\sqrt{(2j-j'+m'+1)(j'-m')(j+m+1)(j-m)} \nonumber \\
\label{Eq:fjA'}
& & \times
\text{Re} [ \langle j'; m' | \hat{A}_{j'} | j ; m \rangle
(\langle j' ; m'+1 | \hat{A}_{j'} | j ; m+1 \rangle )^{\ast} ]
\end{eqnarray}
The second term of the above formula majorizes the second term
of Eq.~(\ref{Eq:fj'Aj'}), since for $j>j'$ we have
\begin{equation}
\sqrt{2j-j'+m'+1} \ge \sqrt{j'+m'+1}
\end{equation}
and all the other factors have been assumed to be nonnegative.
This observation can be combined with the decomposition of the first
term in Eq.~(\ref{Eq:fjA'}) into two parts proportional
to $m'$ and $j-j'$,
which yields:
\begin{equation}
f_j(\hat{A'}) \ge f_{j'} (\hat{A}_{j'}) + (j-j') g(\hat{A}_{j'}).
\end{equation}
Since $j>j'$ and we have assumed that $g(\hat{A}_{j'}) \ge 0$,
this proves that Eq.~(\ref{Eq:Condforf}) is indeed satisfied. Of
course, the condition $g(\hat{A}_{j'}) \ge 0$ is fulfilled automatically
for all the operators relevant to the trade-off,
as according to our discussion from Sec.~\ref{Sec:Decomposition} the
whole region of interest for $g$ is $0 \le g \le j$.

\section{Optimization}
\label{Sec:Optimization}

We will now show that the search for the trade-off curve can be decomposed
into a set of even simpler independent constrained optimization problems.
To proceed further, it will be convenient to introduce vector
notation. Let us define:
\begin{eqnarray}
l_k & = & - j + \max(0,k) \nonumber \\
u_k & = & j + \min(k,0)
\end{eqnarray}
where the index $k$ is from the range $-2j \le k \le 2j$.
For brevity, we also denote:
\begin{equation}
a^k_{m} = \langle j ; m - k | \hat{A}_{j} | j ; m \rangle
\end{equation}
where $ l_k \le m \le u_k $, and we assume that all the matrix elements are
real and nonnegative. We can now introduce $4j+1$ real vectors:
\begin{equation}
{\bf a}_{k} = ( a^{k}_{l_k}, a^{k}_{l_k+1}, \ldots , a^{k}_{u_k}).
\end{equation}
The length of the vector with the index $k$ is equal to $2j+1-|k|$.
These vectors are diagonal stripes of the matrix $\langle j; m |
\hat{A}_{j} | j ; n \rangle$.
Using the vector notation, we have:
\begin{mathletters}
\begin{eqnarray}
f & = & \sum_{k=-2j}^{2j} f^k ({\bf a}^{k}) \\
g & = & \sum_{k=-2j}^{2j} g^{k} ({\bf a}^{k}) \\
1 & = & \sum_{k=-2j}^{2j} h^{k} ({\bf a}^{k})
\end{eqnarray}
\end{mathletters}
with
\begin{mathletters}
\begin{eqnarray}
\label{Eq:frbfar}
f^{k} ({\bf a}^{k}) & = & \sum_{m=l_k}^{u_k} m(m-k) (a_m^{k})^{2}
+ \sum_{m=l_k}^{u_k-1} \gamma_m^k a_m^k a_{m+1}^{k}
\\
g^{k} ({\bf a}^{k}) & = & \sum_{m=l_k}^{u_k} m (a_m^{k})^{2}
\\
h^{k} ({\bf a}^{k}) & = & \sum_{m=l_k}^{u_k} (a_m^{k})^{2}
\end{eqnarray}
\end{mathletters}
and
\begin{equation}
\label{Eq:gammakm}
\gamma_m^{k} =
\sqrt{(j-m)(j+m+1)(j+k-m)(j-k+m+1)}.
\end{equation}
We can now use the same reasoning as before to restrict our interest
to a single vector ${\bf a}^k$ with a fixed value of $k$.  This vector
should be normalized to unity, $({\bf a}^{k})^{T} {\bf a}^{k} = 1$.
The allowed region for $f$ and $g$ is defined as a union of regions
bounded by curves obtained for different $k$'s, with the index $k$
running from $-2j$ to $2j$. We will now discuss several properties of
the curves depending on the sign of $k$, which will allow us to restrict
our search for the optimality curve to a smaller set of $k$'s.

\subsection{Case $k < 0$}

For negative $k$, we note that given an arbitrary vector ${\bf a}^{k}$,
one can use its elements in the same order to construct a vector
${\bf a}^{-k}$ simply by taking ${\bf a}^{-k} = {\bf a}^{k}$.
A simple calculation shows that:
\begin{eqnarray}
f^{-k} ({\bf a}^{-k} ) & = & f^{k} ({\bf a}^{k}) \nonumber \\
g^{-k} ({\bf a}^{-k} ) & = & g^{k} ({\bf a}^{k}) - k
\end{eqnarray}
Thus, one can obtain from any trade-off curve for $k<0$ a certain
trade-off curve for $-k > 0$ which is shifted along the $g$ axis towards
higher values. All the trade-off curves for $k<0$ are hence suboptimal,
and we can further restrict our attention only to the case of $ k \ge 0$.

\subsection{Case $k \ge 0$}
\label{SubSec:kge0}

We will now show that the trade-off curves obtained for $k$'s greater
or equal to $\sqrt{2j}$ lie completely within the region bounded by
the curve corresponding to $k=0$. This will allow us to exclude
all $k \ge \sqrt{2j}$ from further analysis. In order to prove the above
lemma, we will demonstrate that the maximum value of $f$ attained
by the trade-off curves for $k \ge \sqrt{2j}$ lies below the minimum
value of $f$ on the trade-off curve obtained for $k=0$.

We start from the observation that the complete trade-off curve for $k=0$
lies above the value $f=j^2$. Indeed, let us define the vector ${\bf
a}^{0} = (\sin\chi,0,\ldots,0,\cos\chi)$ with $\chi$ from the range $0$ to
$\pi/4$. It is straightforward to check that we have $f^{0}({\bf a}^{0})
\ge j^2$ over this range of $\chi$, whereas $g^{0} ({\bf a}^{0}) = j
\cos 2\chi$, which can assume any value between $0$ and $j$.  Thus for
any $g$ from the range $0 \le g \le j$ relevant to the trade-off curve
we have a vector such that the corresponding value of $f$ is larger or
equal to $j^2$. Consequently, the complete trade-off curve occupies the
region of the $(f,g)$ plane defined by the condition $f \ge j^2$.

Next, we prove in Appendix~\ref{Appendix:BoundOnfkbfak} that for an
arbitrary normalized vector ${\bf a}^{k}$, the function 
$f^k ({\bf a}^{k})$ is bounded by:
\begin{equation}
\label{Eq:Boundonfkak}
f^k ({\bf a}^{k}) \le j(j+1) - \frac{k^2}{2}.
\end{equation}
The right-hand side of the above bound can be compared with the
the minimum value of $f^0 = j^2$ on the trade-off curve for $k=0$. 
If for a given $k$ the general upper bound on $f^k$ given
by the right-hand side of
Eq.~(\ref{Eq:Boundonfkak}) is below the value $j^2$, then the trade-off
curve obtained for this specific $k$ will definitely be majorized by
the trade-off curve corresponding to $k=0$ over the whole region of
interest. Hence we can exclude all $k$'s satisfying
$j(j+1) - k^2/2 \le j^2$, which after
simplification yields $k \ge \sqrt{2j}$.
Consequently, we can restrict our attention to nonnegative $k$'s
from the range:
\begin{equation}
\label{Eq:Condonk}
0 \le k < \sqrt{2j}.
\end{equation}
Recalling that $2j=N$, it is thus sufficient to consider $k$'s from
a finite set of only $\lceil \sqrt{N} \rceil$ values. The number
of independent real variables that have to be optimized for a given
$k$ is equal to $N+1-k$.

\section{Numerical procedure}
\label{Sec:Numerical}

Our task is now reduced to finding the trade-off curves for a set
of $k$'s defined in Eq.~(\ref{Eq:Condonk}). To complete this task,
we shall resort to numerical means. For a given $k$, define
$(u_k - l_k +1) \times (u_k -l_k +1)$ real symmetric matrices
${\cal F}^{k}$ and ${\cal G}^{k}$: the matrix ${\cal F}^{k}$
has $l_k(l_k-k), (l_k+1)(l_k-k+1), \ldots , u_k (u_k -k)$ on
the diagonal, and $\gamma^{k}_{l_k}/2, \gamma^{k}_{l_k+1},
\ldots , \gamma^{k}_{u_r -1}$ on either side of the diagonal.
The matrix ${\cal G}^{k}$ has $l_k, l_k +1, \ldots , u_{k}$
on the diagonal. We can now use the method of Lagrange multipliers
to find the maximum of $f$ having fixed the value of $g$. Specifically,
we need to optimize the expression
\begin{equation}
\label{Eq:MaximizeThis}
f^{k} ({\bf a}^{k} )
+ \lambda g^{k} ({\bf a}^{k} )
- \mu h^{k} ({\bf a}^{k} )
=
({\bf a}^{k})^{T} ( {\cal F}^{k} + \lambda {\cal G}^{k} - \mu \openone )
{\bf a}^{k}
\end{equation}
with the constraints:
\begin{eqnarray}
g^{k} ({\bf a}^{k}) & = & g \nonumber \\
h^{k} ({\bf a}^{k}) & = & 1
\end{eqnarray}
and $\lambda, \mu$ being the Lagrange multipliers.

Differentiating the right hand side of Eq.~(\ref{Eq:MaximizeThis})
with respect to the elements of
the vector ${\bf a}^{k}$ we obtain that the maximum occurs when
\begin{equation}
( {\cal F}^{k}+ \lambda {\cal G}^{k} - \mu I ) {\bf a}^{k}
=
{\bf 0},
\end{equation}
that is, ${\bf a}^{k}$ is an eigenvector of the matrix ${\cal F}^{k}+
\lambda {\cal G}^{k}$ corresponding to its maximum eigenvalue. Assuming
that this eigenvector is normalized to one, the value of $g$ is given by
the product $({\bf a}^{k})^{T} {\cal G}^{k} {\bf a}^{k}$, which implicitly
depends on $\lambda$ through the vector ${\bf a}^{k}$. In order to plot
the trade-off curve as a function $f^{k}(g^{k})$ we would need to invert
this relation. However, we can equivalently consider the trade-off curve
as parameterized by the Lagrange multiplier $\lambda$ running from $0$
to $\infty$.  The value $\lambda = 0$ corresponds to optimizing $f^k$
without any constraint on $g^k$, whereas in the limit $\lambda \rightarrow
\infty$ the optimization gives the maximum attainable value of $g^k$
and the corresponding value of $f^k$.  Thus the procedure of calculating
the trade-off curve for a given $k$ can be summarized as follows: for
$\lambda$'s from the range $0 \le \lambda < \infty$ find the normalized
eigenvector of ${\cal F}^{k}+ \lambda {\cal G}^{k}$ corresponding to
the largest eigenvalue, and use it to calculate $({\bf a}^{k})^{T} {\cal
F}^{k} {\bf a}^{k}$ and $({\bf a}^{k})^{T} {\cal G}^{k} {\bf a}^{k}$ which
give the point $(f^k(\lambda), g^{k}(\lambda))$ on the trade-off curve.

\begin{figure}

\vspace*{\fill}

\begin{center}
\epsfig{file=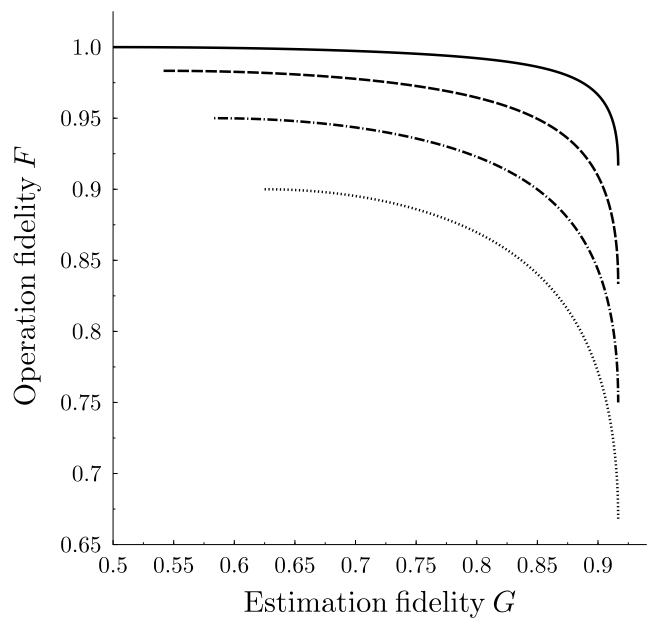,width=4in}
\end{center}

\vspace*{\fill}

\caption{The trade-off curves for the ensemble of $N=10$ qubits obtained
by solving the optimization problem for $k=0$ (solid), $k=1$ (dashed),
$k=2$ (dotted-dashed), and $k=3$ (dotted).
The allowed region is a union of the regions
bounded by each of these curves.}
\label{Fig:FixedN}
\end{figure}

We have performed numerically the task of finding the eigenvectors
${\bf a}^{k}$. In numerical calculations, it is convenient to change the
parameterization of the trade-off curve according to $\lambda = x/(1-x)$
with $x$ running from $0$ to $1$, and to diagonalize the rescaled
matrix $(1-x) {\cal F}^{k} + x {\cal G}^{k}$.  We note that this matrix
is of the tridiagonal form, for which there exist efficient numerical
algorithms\cite{PressTriDiag}. After finding the normalized eigenvectors
we can plot the parameterized trade-off curve $(f^k(x),g^k(x))$ for
all relevant $k$'s from the range given by Eq.~(\ref{Eq:Condonk}),
and next find on this basis the region of fidelities allowed in quantum
mechanics. To illustrate this procedure, we plot in Fig.~\ref{Fig:FixedN}
the trade-off curves for $N=10$ and the relevant range of $k$'s. It
is seen that the quantum mechanically allowed region is bounded by a
single trade-off curve obtained for $k=0$. This was the case also in
all other cases we investigated numerically. The actual bounds on the
operation fidelity $F$ versus the estimation fidelity $G$ for several
values of $N$ are shown in Fig.~\ref{Fig:Results}.

On the basis of our numerical studies we conjecture that the trade-off
curve for $k=0$ is always optimal. This conjecture is supported by the
fact that both the extreme points corresponding to $x=0$ and $x=1$ are
attained only for $k=0$. Indeed, we have seen in Sec.~\ref{SubSec:kge0}
that for a given nonnegative $k$ the whole trade-off curve lies below
$j(j+1) - k^2/2$. Consequently, the maximum value of $f = j(j+1)$
corresponding to unit operation fidelity is achieved only for $k=0$.
This is also the case of the other extreme point corresponding to
the optimal estimation: it is straightforward to note that
for any $k \ge 0$ the expression for $g^k$ has the same maximum value
equal to $g=j$. This value is obtained for the unique vector
of the form
${\bf a}^{k} = (0,0,\ldots , 1)$. The corresponding value
of $f^k$ is $f^k=j(j-k)$. Thus, for optimized $g$ the largest attainable
value of $f$, equal to $j^2$, is obtained only for $k=0$.

\section{Attainability of the bound}
\label{Sec:Attainability}

We will now show that the trade-off curves computed in the previous
section are tight. i.e.\ that they can be attained by physically
realizable operations. So far, we have considered only the trade-off
curve generated by a single operator
$\hat{A} :{\cal H}_{j} \rightarrow {\cal H}^{\otimes N}$ satisfying
the condition $\text{Tr} ( \hat{A}^{\dagger} \hat{A} ) = 1$. The critical
step which allowed us to focus on a single operator was the replacement
of the full trace-preserving condition in Eq.~(\ref{Eq:Trpresj}) by its trace.
We will now present a method for constructing a complete quantum
operation from a single operator $\hat{A}$ such that it generates the same
point on the fidelity trade-off curve.

The classical outcome of the operation we construct 
has the continuous
form of a triplet of Euler angles which we will denote by $\Xi$.
The operation element corresponding to a specific outcome 
$\Xi$ is given by:
\begin{equation}
\hat{A}_{\Xi} = \sqrt{2j+1} \hat{U}(\Xi) \hat{A} \hat{U}^\dagger
(\Xi)
\end{equation}
It is straightforward to verify that the operation fidelity $F$
and the estimation fidelity $G$ for this operation are given respectively
by:
\begin{equation}
F = \frac{1}{2} + \frac{1}{2j(j+1)} f(\hat{A})
\end{equation}
and
\begin{equation}
G = \frac{1}{2} + \frac{1}{2(j+1)} g (\hat{A})
\end{equation}
where $f(\hat{A})$ and $g (\hat{A})$ are defined in Eqs.~(\ref{Eq:fAdef})
and (\ref{Eq:gAdef}). This confirms that we obtain the same point on
the trade-off curve as for the operator $\hat{A}$ itself. The
only condition we need to check is the completeness of the operation
on the fully symmetric subspace ${\cal H}_j$:
\begin{equation}
\label{Eq:AXiCompleteness}
\int d\Xi \, \hat{A}^{\dagger}_{\Xi} \hat{A}_{\Xi}
= \hat{\openone}_j.
\end{equation}
Of course, outside ${\cal H}_{j}$ the value of this integral vanishes,
as the operator $\hat{A}$ is assumed to be zero there and the
rotation matrices $\hat{U}(\Xi)$ do not mix
subspaces with different values of the angular momentum.

In order to prove that the completeness condition
given by Eq.~(\ref{Eq:AXiCompleteness}) is indeed satisfied, let
us consider the matrix element of the left-hand side of the above expression
in the eigenbasis of the angular momentum operator $\hat{J}^{z}_{j}$:
\begin{eqnarray}
\lefteqn{\int d\Xi \, \langle j ; m |
\hat{A}^\dagger_{\Xi} \hat{A}_{\Xi} | j ; n \rangle }
\nonumber \\
& = & (2j+1) \sum_{m',n' =-j}^{j} \int d\Xi
\langle j ; m | \hat{U} (\Xi) | j;m' \rangle
\langle j;m' | \hat{A}^\dagger \hat{A} |j ; n' \rangle
\langle j ; n' | \hat{U}^{\dagger} ( \Xi ) | j ; n \rangle .
\end{eqnarray}
The integral over the product of the elements of rotation matrices
$\langle j ; m | \hat{U} (\Xi) | j;m' \rangle =
{\cal D}^{j}_{mm'}(\Xi)$ can be performed explicitly using
the standard formula:
\begin{equation}
\int d\Xi \,
{\cal D}^{j}_{mm'} (\Xi) [ {\cal D}^{j}_{nn'} (\Xi ) ]^\ast
=
\frac{\delta_{mn} \delta_{m'n'}}{2j+1}.
\end{equation}
With the help of the above identity we have:
\begin{eqnarray}
\lefteqn{\int d\Xi \, \langle j ; m |
\hat{A}^\dagger_{\Xi} \hat{A}_{\Xi} | j ; n \rangle }
\nonumber \\
& = & (2j+1) \sum_{m',n' =-j}^{j} \int d\Xi \,
{\cal D}^{j}_{mm'} (\Xi) [ {\cal D}^{j}_{nn'} (\Xi ) ]^\ast
\langle j;m' | \hat{A}^\dagger \hat{A} |j ; n' \rangle
\nonumber \\
& = & \sum_{m',n'=-j}^{j} \delta_{mn} \delta_{m'n'}
\langle j ; m' | \hat{A}^\dagger \hat{A} | j ; n' \rangle
= \delta_{mn} \text{Tr} (\hat{A}^\dagger \hat{A} ) = \delta_{mn}
\end{eqnarray}
This completes the proof that the operation $\{\hat{A}_{\Xi}\}$
satisfies the full trace-preserving condition in the
symmetric subspace of the $N$-qubit Hilbert space. Consequently,
the trade-off curve calculated for a single operator $\hat{A}$ is
attained also by complete quantum operations.

\section{Discussion}
\label{Sec:Discussion}

In this paper we calculated the fidelity trade-off for finite ensembles
of identically prepared qubits. This trade-off relates the quality of
estimating the quantum state of the qubits to the minimum disturbance of
the original that has to be introduced in course of this procedure. The
obtained trade-off curves can also be viewed as a characterization
of a specific asymmetric quantum cloning procedure, which given $N$
qubits produces the same number of clones with a decreased fidelity $F$,
and additionally an arbitrarily large number of clones with a lower
fidelity $G$.

The calculation of the trade-off curve was based on a combination of
analytical techniques and numerical calculations. The results obtained
analytically allowed us to reduce significantly the complexity of
the optimization problem. One should note that a single operator
$\hat{A} : {\cal H}_j \rightarrow {\cal H}^{\otimes N}$ mapping the
fully symmetric subspace onto the whole Hilbert space of $N$ qubits
contains $2^{N+1}(N+1)$ independent real variables.  If one wanted to
find numerically the trade-off curve assuming such general form of the
operator $\hat{A}$, the number of parameters in the optimization problem
would explode exponentially with the size of the ensemble.  Fortunately we
were able to demonstrate that the problem of finding the trade-off curve
can be reduced to $\lceil \sqrt{N} \rceil$ independent optimization
problems, each involving only no more than $N+1$ real parameters.
This is a substantial reduction of the problem, which enables one to
handle numerically much larger ensembles of qubits.

There are several elements of our work which could be investigated
further.  First, it would be interesting to prove our conjecture that the
trade-off curve obtained for $k=0$ is always optimal.  This would reduce
further the complexity of the problem remaining to solve numerically. We
have made several observations that might be helpful in this proof. First,
numerical calculations suggest that the eigenvalues of the matrix ${\cal
F}^{k}$ considered in Sec.~\ref{Sec:Numerical} belong to the analytically
defined set $\{ -\nu (\nu -1)/2 + 2j\nu - j^2 | \nu = 0,1,\ldots, u_k-l_k
\}$. The largest of these eigenvalues is $j(j+1) - k(k+1)/2$, which itself
improves the upper bound given in Eq.~(\ref{Eq:Boundonfkak}). Inspection
of numerical results suggests also that the value of $g^{k}({\bf a}^{k})$
corresponding to maximized $f^{k}({\bf a}^{k})$ is equal to $k/2$. Thus,
both the extreme points of all trade-off curves for $k>0$ lie beneath the
one obtained for $k=0$. This observation, combined with a demonstration
that the curves have appropriate monotonicity and convexity properties,
might prove the universally optimal character of the $k=0$ curve.

Another interesting direction is investigating in more detail quantum
operations that saturate the trade-off inequality. We have shown that
given a single operator which generates the values of $F$ and $G$ lying
on the trade-off curve, one can construct a complete quantum operation
that satisfies the full trace-preserving condition.  The described
operation had a continuous classical outcome in the form of a triplet of
Euler angles. It would be interesting to investigate operations with a
finite \cite{DerkBuzePRL98} (and possibly minimal \cite{LatoPascPRL98})
number of outcomes that also saturate the quantum mechanical bound on
the fidelities.

\section*{Acknowledgements}

We wish to acknowledge useful discussions with T. Berger, C. M. Caves,
and C. A. Fuchs. This research was supported by ARO-administered MURI
Grant No.\ DAAG-19-99-1-0125 and the European Union Project EQUIP
(contract IST-1999-11053).

\appendix

\section{Multiplicities of angular momentum representations}
\label{Appendix:Multiplicities}

In this Appendix, we derive the mutliplicities $\mu_{j'}$ of
subspaces with the fixed value of the angular momentum $j'$ appearing
in the decomposition of the Hilbert space of $N$ qubits.
Let us consider the operator:
\begin{equation}
\hat{Z}(\beta) = \bigotimes_{i=1}^{N} \exp(\beta \hat{\sigma}^{z}_{i})
=
\exp (2\beta \hat{J}^{z} ),
\end{equation}
where the subscript $i$ enumerates the qubits.
From the tensor-product representation given on the
left-hand side of the above formula we immediately obtain that
\begin{equation}
\label{Eq:TrZbeta}
\text{Tr} [ \hat{Z} (\beta) ] = (2 \cosh \beta)^N.
\end{equation}
On the other hand, summation of the trace of the operator
$\exp(2\beta \hat{J}^{z})$ in all the subspaces yields:
\begin{equation}
\text{Tr} [\hat{Z}(\beta)] = \sum_{j' = j - \lfloor j \rfloor}^{j}
\mu_{j'} \sum_{m=-j'}^{j'} e^{2 m \beta } =
\sum_{j'= j- \lfloor j \rfloor}^{j} \mu_{j'}
\frac{\sinh \beta(2j'+1)}{\sinh\beta}.
\end{equation}
Comparing equal powers of $e^\beta$ in this expression with the
expansion of the left-hand side of Eq.~(\ref{Eq:TrZbeta}) 
given by $(2\cosh\beta)^N$ yields:
\begin{equation}
\mu_{j'} = { 2j \choose {j + j'}} - {2j \choose {j + j' +1}}
= \frac{2j'+1}{2j+1} { {2j+1} \choose {j-j'}}.
\end{equation}
It is seen that the total number of subspaces in the decomposition
(\ref{Eq:HDecomposition}) is $\sum_{j' = j - \lfloor j \rfloor}^{j}
\mu_{j'} = { 2j \choose {\lfloor j \rfloor}}$.

\section{Evaluation of integrals $\hat{K}_{\tau}$}
\label{Appendix:K}

In this Appendix we calculate explicitly the integrals $\hat{K}_{\tau}$
defined in Eq.~(\ref{Eq:Kdef}). For this purpose, 
it is convenient to switch to the angular
momentum representation resulting from the decomposition of the tensor
product Hilbert
space of the $N$ qubits into the direct sum of subspaces with the fixed
value of the total angular momentum operator.
In this representation, the state
$|\Omega\rangle^{\otimes N}$ lies in the completely symmetric
subspace characterized by the angular momentum $j=N/2$ and it is given by 
\begin{equation}
|\Omega \rangle^{\otimes N} = \hat{U} (\Omega)
|j;j\rangle,
\end{equation}
where $|j;j\rangle$ is the eigenvector of $\hat{J}^z$ corresponding
to the eigenvalue $j$, and $\hat{U}(\Omega)$ is the rotation matrix
for the angular momentum $j$. Of course, rotations 
rotations cannot transfer the state $|j;j\rangle$ beyond the fully
symmetric subspace. Consequently, the operators $\hat{K}_{\tau}$ are nonzero
only in this subspace. In the basis of the eigenvectors of the
operator $\hat{J}^z$, the matrix elements of these operators
are given by:
\begin{equation}
\langle j;m | \hat{K}_{\tau} | j ; n \rangle 
=
\int d\Omega \, k_\tau (\Omega) \langle j;m | \hat{U} (\Omega)
| j; j \rangle
\langle j ; j | \hat{U}^{\dagger} (\Omega) | j ; n \rangle
\end{equation}
We shall use the standard notation from Ref.~\cite{Brink} to denote the
matrix elements of the unitary rotation operators appearing
in the above formula:
\begin{eqnarray}
\langle j ; m | \hat{U}(\Omega) | j ; j \rangle
& = & {\cal D}^{j}_{mj}(\Omega)
\nonumber \\
\langle j ; j | \hat{U}^\dagger(\Omega) | j; n \rangle
& = & [ {\cal D}^{j}_{nj}(\Omega)]^{\ast} = 
(-1)^{n-j} {\cal D}^{j}_{-n \, -j} (\Omega),
\end{eqnarray}
where in the second line we have made use of the symmetry properties
of the rotation matrix elements.

The functions $k_\tau(\Omega)$ can also be expressed as elements of
rotation matrices for the value of the total angular momentum equal to one:
\begin{eqnarray}
k_{-1}(\Omega) & = & \sqrt{2} {\cal D}^{1}_{-1 0}(\Omega)
\nonumber \\
k_{0}(\Omega) & = & {\cal D}^{1}_{00}(\Omega)
\nonumber \\ 
k_{1}(\Omega) & = & -\sqrt{2} {\cal D}^{1}_{10}(\Omega)
\end{eqnarray}
With this notation, we can use the standard expression for the integrals
of triple products of rotation matrix elements in terms of the Wigner
3-$j$ symbols:
\begin{eqnarray}
\langle j ; m | \hat{K}_{\pm 1} | j ; n \rangle & = &
\mp \sqrt{2} (-1)^{n-j} \int d\Omega \, 
{\cal D}^{j}_{mj}(\Omega) {\cal D}^{j}_{-n\,-j} (\Omega)
{\cal D}^{1}_{\pm 10}(\Omega)
\nonumber \\
& = & \sqrt{2} (-1)^{n-j}
\left(
\begin{array}{ccc}
j & j & 1 \\
m & -n & \pm 1
\end{array}
\right)
\left(
\begin{array}{ccc}
j & j & 1 \\
j & -j & 0
\end{array}
\right)
\end{eqnarray}
and
\begin{eqnarray}
\langle j ; m | \hat{K}_{0} | j ; n \rangle
& = &
(-1)^{n-j} \int d\Omega \, 
{\cal D}_{mj}^{j} (\Omega) {\cal D}_{-n\,-j}^{j} (\Omega)
{\cal D}_{00}^{1} (\Omega)
\nonumber \\
& = &
(-1)^{n-j}
\left( \begin{array}{ccc}
j & j & 1 \\
m & -n & 0
\end{array}
\right)
\left(
\begin{array}{ccc}
j & j & 1 \\
j & -j & 0 
\end{array}
\right).
\end{eqnarray}
Inserting the explicit form of the 3-$j$ symbols yields:
\begin{equation}
\langle j ; m | \hat{K}_{\pm 1} | j ; n \rangle 
= \delta_{m \pm 1,n} \frac{\sqrt{(j \pm n)(j \mp n+1)}}{(j+1)(2j+1)}
\end{equation}
and
\begin{equation}
\langle j ; m | \hat{K}_{0} | j ; n \rangle 
 = 
\delta_{mn} \frac{m}{(j+1)(2j+1)}
\end{equation}
The expressions appearing on the right hand sides of the above formulas
are proportional to the operators $\hat{J}^{\mp}$ and $\hat{J}^z$ restricted
to the completely symmetric subspace with the total angular momentum $j$.
Thus, the integrals $\hat{K}_{\tau}$ can be conveniently written as
\begin{equation}
\hat{K}_{\pm 1} = \frac{1}{(j+1)(2j+1)} \hat{J}^{\mp}_{j}
\end{equation}
and
\begin{equation}
\hat{K}_{0} = \frac{1}{(j+1)(2j+1)} \hat{J}^{z}_{j}
\end{equation}
where by the subscript $j$ we have denoted the angular momentum operators
truncated to the subspace characterized by the angular momentum $j$.

\section{Derivation of the bound on
\mbox{\lowercase{$f^{k} ({\bf a}^k)$}}}
\label{Appendix:BoundOnfkbfak}

Here derive
an upper bound on the functions $f^k({\bf a}^k)$,
defined in Eq.~(\ref{Eq:frbfar}), for $k \ge 0$. For this purpose,
let us rewrite $\gamma^{k}_{m}$ defined in Eq.~(\ref{Eq:gammakm})
to the form:
\begin{equation}
\gamma^{k}_{m} =
\sqrt{
\left[
\left( j - m + \frac{k}{2} \right)^2 - \left( \frac{k}{2} \right)^2
\right]
\left[
\left( j + m + 1 - \frac{k}{2} \right)^2 - \left( \frac{k}{2} \right)^2
\right]
}
\end{equation}
Using the inequality:
\begin{equation}
\sqrt{ (x_1^2 - x_2^2)(x_3^2 - x_4^2)} \le x_1 x_3 - x_2 x_4
\end{equation}
valid for any real $x_1 \ge x_2 \ge 0$ and $x_3 \ge x_4 \ge 0$,
we obtain that 
\begin{equation}
\label{Eq:gammakmle}
\gamma^{k}_{m} \le \delta^{k}_{m},
\end{equation}
where:
\begin{equation}
\label{Eq:deltakmle}
\delta^{k}_{m} \le j(j+1) - \left(m - \frac{k}{2} \right)
\left(m - \frac{k}{2} + 1 \right) - \left( \frac{k}{2} \right)^2
\end{equation}
We will also use the inequality between the geometric and arithmetic
means to estimate:
\begin{equation}
\label{Eq:akmakm+1le}
a^{k}_{m} a^{k}_{m+1} \le \frac{1}{2}[ (a^{k}_{m})^2 + (a^{k}_{m+1})^2].
\end{equation}
We can now use the inequalities given in Eqs.~(\ref{Eq:gammakmle})
and (\ref{Eq:akmakm+1le}) to find an upper bound on the second term
in Eq.~(\ref{Eq:frbfar}). A simple rearrangement of the terms yields:
\begin{eqnarray}
f^{k} ({\bf a}^{k}) & \le & \sum_{m=l_k}^{u_k} m(m-k) (a^{k}_{m})^2 +
\frac{1}{2} \sum_{m=l_k}^{u_k-1} \delta^{k}_{m}
[ (a^{k}_{m})^2 + (a^{k}_{m+1})^2] 
\nonumber \\
 & = & 
\sum_{m=l_k}^{u_k} m(m-k) (a^{k}_{m})^2 
+ \frac{1}{2} \delta_{l_k}^{k} ( a_{l_k}^{k} )^2
+ \frac{1}{2} \delta_{u_k-1}^{k} (a_{u_k}^{k})^2
+ \sum_{m=l_k+1}^{u_k-1} \frac{1}{2} (\delta_{m-1}^{k}
+ \delta_{m}^{k}) (a_m^k)^2.
\nonumber \\
& &
\end{eqnarray}
We can now add to the right-hand side two terms of the form
$\delta_{l_k-1}^{k} (a_{l_k}^{k})^2/2$
and $\delta_{u_k}^{k} (a_{u_k}^{k})^2/2$ which are nonnegative
as can be easily checked, and use the identity:
\begin{equation}
\frac{1}{2}(\delta_{m-1}^{k} + \delta_{m}^{k} )
=
j(j+1) - m(m-k) - \frac{k^2}{2}.
\end{equation}
This finally yields:
\begin{eqnarray}
f^k ({\bf a}^{k}) & \le & \sum_{m=l_k}^{u_k} m(m-k) (a^{k}_{m})^2 +
 \sum_{m=l_k}^{u_k} \left( j(j+1) - m(m-k) -
 \frac{k^2}{2}  \right) (a^{k}_{m})^2
 \nonumber \\
 \label{Eq:boundonfk}
 & = & \left( j(j+1) - \frac{k^2}{2} \right) \sum_{m=l_k}^{u_k}
( a_m^k )^2.
\end{eqnarray}
If the vector ${\bf a}^{k}$ is normalized to unity, i.e.\
$\sum_{m=l_k}^{u_k} ( a_m^k )^2 = 1$, we obtain the bound
given in Eq.~(\ref{Eq:Boundonfkak}).

We note that for $k=0$ the derived bound is tight as it can be achieved
for ${\bf a}^{k=0} = (1/\sqrt{2j+1}, \ldots , 1/\sqrt{2j+1})$.

\end{document}